\documentclass[aps,prd,superscriptaddress,10pt,showpacs,notitlepage,  
nofootinbib
]{revtex4-1}
\usepackage{bm}
\usepackage{amsmath,amssymb,amsthm}
\usepackage{latexsym,graphicx,color,subfigure}
\usepackage{enumerate}
\usepackage{amssymb}
\usepackage{hyperref}
\usepackage{mathtools}
\usepackage{graphicx}

\usepackage{array} 
\newcolumntype{P}[1]{>{\centering\arraybackslash}p{#1}}

\graphicspath{ {images/} }

\newcommand{\be}{\begin{equation}}
\newcommand{\ee}{\end{equation}} 
\newcommand{\beq}{\begin{eqnarray}}
\newcommand{\eeq}{\end{eqnarray}}

\newcommand{\bea}{\begin{eqnarray}}
\newcommand{\eea}{\end{eqnarray}}

\def\tc{\textcolor{red}}
\def\tc2{\textcolor{blue}}

\allowdisplaybreaks[3]
\def\simge{\mathrel{
   \rlap{\raise 0.511ex \hbox{$>$}}{\lower 0.511ex \hbox{$\sim$}}}}
\def\simle{\mathrel{
   \rlap{\raise 0.511ex \hbox{$<$}}{\lower 0.511ex \hbox{$\sim$}}}}
\def\bigs{\mathrel{
   \rlap{\raise 0.531ex \hbox{$>$}}{\lower 0.531ex \hbox{$<$}}}}

\usepackage[normalem]{ulem}

\renewcommand\sout{\bgroup \color{blue} \ULdepth=-.5ex \ULset}

\begin{document}
	\preprint{APS/123-QED}
	
	\title{Masses and decays of the bottom-charm hybrid meson $c \bar{b} g$}
	\author{T. Miyamoto}
	\email{miyamoto-tomokazu-gv@ynu.ac.jp, \\ tomokazu.miyamoto10@physics.org \\ tomokazu.miyamoto2@keio.jp}
	\affiliation{Graduate School of Engineering Science, Yokohama National University, Tokiwadai 79-1, Yokohama, Kanagawa 240-8501, Japan}	
	\affiliation{Research and Education Center for Natural Sciences, Keio University, Hiyoshi 4-1-1, Yokohama, Kanagawa 223-8521, Japan}

	\author{S. Yasui}
	\email{yasuis@keio.jp}
	\affiliation{Research and Education Center for Natural Sciences, Keio University, Hiyoshi 4-1-1, Yokohama, Kanagawa 223-8521, Japan}

	\date{\today} 
	\thispagestyle{empty}
	\begin{abstract}
	We study gluonic excitations inside a $B_{c} $ meson in the constituent gluon model, treating a bottom-charm hybrid meson $ c \bar{b} g $ as a three-body system. We obtain the mass spectra for the hybrid mesons with magnetic gluon and electric gluon and see that their lowest states appear above the $ D B $ threshold. Also, we consider the decays of the low-lying states of the hybrid meson into $ DB , \; D^{*} B, DB^{*} $, $ D^{*}B^{*} $, $ D_{s} B_{s} $, and $D_{s} B_{s}^{*} $ mesons, developing an existing model for the strong decays. We estimate their partial decay widths and find that the widths have a heavy dependence on the final meson states. We argue that the accuracy of our model will be tested experimentally when the branching ratios of the decays are measured. Our results suggest that there could be a prospect that the hybrid meson will be discovered as its first excited state rather than its lowest state.  
	
	 	
	\end{abstract}
	
	\maketitle
	
	\onecolumngrid 
	

  \section{Introduction}

  It has been argued for a long time that the experimental verification of exotic hadrons will have far-reaching implications for quantum chromodynamics (QCD), not just because it will give the quantum field theory of the strong interaction another credit for the theory's accuracy, but because it will make QCD researchers search for nonperturbative properties of the theory with renewed enthusiasm. While it is still difficult to experimentally confirm the existence of exotic hadrons, indirect evidence of it has been accumulating gradually. After the dark ages of exotic hadron research, it finally blossomed into a growth area of research by virtue of the discovery of X(3872) at Belle in 2003 \cite{Choi2003}. The event has been described as a milestone in exotic hadron research; in fact, dozens of exotic hadron candidates have been discovered.
  
  In particular, it is widely known that $ \psi(4260) $ \cite{Tanabashi2018} aka Y(4260) has been a high-profile exotic meson candidate since it was discovered in 2005 \cite{Aubert2005}. The question on the constituents of $ \psi (4260) $ has been considered from many different angles: a hadrocharmonium \cite{Dubynskiy2008}, a $ \bar{D} D_{1}(2420) $ molecule \cite{Wang2013}, a tetraquark, a diquark-antidiquark bound state (diquarkonium) $ [cq][\bar{c}\bar{q}'] $ \cite{Maiani2005_Y4260}, and a hybrid meson $ c \bar{c} g$ \cite{Kou2005}. Although the discussions about $ \psi(4260) $ have not been concluded yet, it has been suggested that the particle might be a multiquark meson or a hybrid quarkonium \cite{Olsen2018}.  

  $ \psi(4260) $ is not the only one which has been considered to be a hybrid meson candidate. For instance, $ \phi(2170) $ has been regarded as a candidate for a strangeonium \cite{Ding2007}, and there could be a possibility that $ \psi(4360) $ is a hybrid charmonium \cite{Ding2008}. It is noteworthy that the decays of $ \psi(4260) $ into $D \bar{D} $ have not been observed yet, despite the fact that the particle sits above the $D \bar{D} $ threshold; this observational fact suggests that it might be a magnetic gluon hybrid \cite{Kou2005}. Likewise, it could be reasonable to expect that $ \psi(2170) $ and $ \psi(4360) $ might be magnetic gluon quarkonia.

  Still, confirming the existence of a hybrid quarkonium is experimentally challenging; the degrees of freedom of a gluon in an ordinary quarkonium are integrated out (and accordingly appear as a potential between two quarks), which makes it difficult to measure gluonic excitations inside the particle. Having said that, the discoveries of hybrid quarkonium candidates have provided useful insights into gluonic excitations inside a quarkonium. In parallel with this, several models have been proposed so far to interpret gluonic excitations inside a meson: the flux tube model \cite{Isgur1985}, the MIT bag model \cite{Barnes1979}, the quark confining string model \cite{Giles1976}, and the constituent gluon model \cite{Horn1978, Yaouanc1985}. In the constituent gluon model, a massless $ J^{P} =1^{-} $ gluon is assumed to interact with quarks via potentials.

  From a wider perspective, a quarkonium is not the only hadron inside which gluonic excitations could appear, given that it is theoretically possible that the phenomena could emerge in other heavy hadrons such as $ B_{c}^{\pm} $ and $ \Omega_{ccc}^{++} $. The purpose of the present study is to expand our horizons by casting light on gluonic excitations inside a $B_{c}^{\pm} $; and thus, we hereafter stick to gluonic excitations inside $ B_{c}^{\pm} $. On the experimental side, there has been no direct evidence that a bottom-charm hybrid meson $ c \bar{b} g $ could exist, in part because detecting the particle is technically more difficult than detecting a hybrid quarkonium. On the theoretical side, we note that this novel particle has not yet been studied extensively. As earlier theoretical studies, we refer to the QCD sum rule analysis and the constituent gluon model. The former showed that several different states of a $ c \bar{b} g $ could exist in the mass range of 6.6 to 8.7 GeV \cite{Chen2014}. In the latter, the mass spectra and the radiative transition widths of the hybrid meson were calculated under the Born-Oppenheimer approximation, which indicated that the lowest states of a $ c \bar{b} g $ appeared in the mass range of 7.4 to 7.6 GeV \cite{Akbar2018}.    

  In the present paper, we adopt the constituent gluon model to discuss whether gluonic excitations could appear inside a $B_{c}$ meson, considering the mass spectra of a $ c \bar{b} g $. We treat a $ c \bar{b} g $ as a three-body system, allowing quarks to move, in contrast to the treatment in Ref.~\cite{Akbar2018} where the distance between charm and antibottom quarks in a $c \bar{b} g $ was fixed. We also discuss the strong decays of a $ c \bar{b} g $ into $ D^{(*)} B^{(*) } $ mesons or $ D^{(*)}_{s} B^{(*)}_{s} $ mesons. As far as we know, the decay widths for these processes are estimated for the first time. We use the notation $D^{(*)}$ to express $D$ or $D^{*} $, and similarly $ B^{(*) } $ for $B$ or $B^{*}$ as well as for $ D^{(*)}_{s}$ and $B^{(*)}_{s} $. It seems to be reasonable that we focus on $J^{P} = 1^{-} $; and it is straightforward to apply the present framework to other quantum numbers.


  This paper is organized as follows. Section II explains our model setting for hybrid mesons, introduces the auxiliary field method for solving the system, and provides the framework of calculating the decay widths for the processes $ c\bar{b} g \rightarrow D^{(*)} B^{(*) } $ or $ D^{(*)}_{s} B^{(*)}_{s} $. Section III presents the results of our numerical calculations and discusses the implications of them. The final section is devoted to our concluding remarks.

  \section{Frameworks for hybrid mesons}
    \subsection{Constituent gluon model for a $c\bar{b}g$ } \label{sec:Hamiltonian_and_auxiliary_field_method}
 
 
    The basic framework of the constituent gluon model for a hybrid meson $ q \bar{q} g $ (where $q$ is a quark) was developed in earlier research \cite{Buisseret2006_hybrids_AF, Buisseret2006_effective_potential_hybrid, Mathieu2009}. Following their lead, and assuming that we can apply the formalism to $ q \bar{q}' g $ (where $q$ and $q'$ are quarks with different flavors), we use a Salpeter-type free Hamiltonian $ H_{0} $ and a potential $ V $ for the constituent particles $q$, $q'$, and $g$ in a $c \bar{b} g$. In relation to this formalism, we use the following physical quantities: the mass $m_{c} $ and momentum $  {\bf p}_{c}  $ of a charm quark $(c ) $, the mass $m_{\bar{b}}$ and momentum ${\bf p}_{\bar{b} } $ of an antibottom quark $(\bar{b}) $, the momentum $ {\bf p}_{g} $ of a constituent gluon $(g)$, the $ c \textrm{-} g $ distance $r_{c g}$, the $ \bar{b} \textrm{-} g $ distance $r_{\bar{b} g }$, and the $ c \textrm{-} \bar{b}    $ distance $  r_{c \bar{b} }$. Using these quantities, we write $ H_{0} $ and $V $ as follows:  
    \begin{align}	
  	H_{0} & = \sqrt{ {\bf p}^{2}_{c} + m^{2}_{c} } + \sqrt{ {\bf p}^{2}_{\bar{b} } + m^{2}_{\bar{b} } } 
  	+ \sqrt{ {\bf p}^{2}_{g }   }  \; ,  
  	\label{eq:Salpeter_type_free_Hamiltonian}
     \\ 
  	V & = \sigma r_{cg} +  \sigma r_{\bar{b} g} + V_{C}    \; , 
    \end{align}
    where $\sigma $ is the string tension for the color confinement and $ V_{C} $ is a color-Coulomb potential which is given by  
     \begin{align}
    	V_{C} = 
    	- \frac{3 \alpha_{s} }{2 r_{cg} }
    	-\frac{3 \alpha_{s} }{2 r_{\bar{b} g}} 
    	+ \frac{\alpha_{s} }{6 r_{c \bar{b} } }   \; , 
    \end{align} 
    with the strong coupling constant $\alpha_{s} $. Here the Casimir factors are taken into account:   
    \begin{align}
    	\frac{1}{4} \langle \lambda_{c} \cdot \lambda_{g}  \rangle = 
    	\frac{1}{4} \langle \lambda_{ \bar{b} } \cdot \lambda_{g}  \rangle =
    	- \frac{3}{2}  \; ,
    	\quad 
    	\frac{1}{4} \langle \lambda_{c} \cdot \lambda_{ \bar{b} }  \rangle = \frac{1}{6}	\;  
    	,
    \end{align}
    where $\lambda_{i}$ ($i=c,\bar{b},g$) is the Gell-Mann matrix in color acting for $c$, $\bar{b}$, and $g$, and the angle brackets mean the expectation value. Then, we have the total Hamiltonian: $ H = H_{0} + V $.

    Once the color-Coulomb potential is introduced, an analytical approach no longer has its advantage over numerical ones. We therefore need to solve the system numerically, but dealing with the (massless) constituent gluon and the Salpeter-type Hamiltonian numerically is difficult. Even if we resolve this problem, there still remains a quantization issue. To address these issues, we use the auxiliary field method. 
    In the present case, employing the method developed in Ref. \cite{Kalashnikova2005}, we introduce the auxiliary fields, $ \mu_{c}, \mu_{\bar{b} } $, and $\mu_{g} $ for $c $, $ \bar{b} $, and $g $, respectively, in order to rewrite the free Hamiltonian \eqref{eq:Salpeter_type_free_Hamiltonian} as follows: 
    \begin{align}
   	H_{0} & = \frac{\mu_{c} + \mu_{\bar{b} } + \mu_{g}  }{2} + 
   	\frac{{\bf p}^{2}_{c} + m^{2}_{c} }{2  \mu_{c} } + 
   	\frac{{\bf p}^{2}_{\bar{b} } + m^{2}_{\bar{b} } }{2 \mu_{\bar{b} } } + 
   	\frac{{\bf p}^{2}_{g} }{2 \mu_{g} }   \; 
   	\nonumber \\
   	& = \frac{\mu_{c} + \mu_{\bar{b} } + \mu_{g}  }{2} + 
   	\frac{ m^{2}_{c} }{2  \mu_{c} } + 
   	\frac{ m^{2}_{\bar{b} } }{2 \mu_{\bar{b} } } + 
   	\frac{ {\bf k}^{2}_{x} }{2 \mu_{x} } + \frac{ {\bf k}^{2}_{y} }{2 \mu_{y} }  \; ,   
    \end{align}  
    where ${\bf k}_{x}$, ${\bf k}_{y}$, $\mu_{x}$, and $\mu_{y  }$ are defined by   
    \begin{align}
    {\bf k}_{x} & = \frac{\mu_{c } {\bf p}_{\bar{b} } - \mu_{\bar{b} } {\bf p}_{c} }{\mu_{c} + \mu_{\bar{b}} } \; , \quad {\bf k}_{y} = {\bf p}_{g} \; 
    ,
    \\         
   	\mu_{x} & = \frac{\mu_{c}  \mu_{\bar{b} } }{\mu_{c} + \mu_{\bar{b}} } \; ,  \quad    
   	\mu_{y} = \frac{  (\mu_{c} + \mu_{\bar{b}}) \mu_{g} }{ (\mu_{c} + \mu_{\bar{b} } )  +  \mu_{g} }  \; . 
    \end{align}
    We then solve the Schr\"{o}dinger-type equations for a hybrid meson: 
    \begin{align} 
    H \Psi = E \Psi  
    \; ,   
    \label{eq:Schrodinger_equation_for_three_body_system}
    \end{align} 
    where $\Psi $ is the wave function for the whole system. In theory, the auxiliary fields are dynamical variables, while in practical calculations we regard them as real numbers in terms of a variational calculation \cite{Kalashnikova2005}. It is known that, under this framework, a calculated value contains roughly a 7 percent numerical error at the maximum \cite{Kalashnikova2005}. As a numerical method for solving Eq.~(\ref{eq:Schrodinger_equation_for_three_body_system}), we mention that we use the hyperspherical formalism \cite{Miyamoto2018}; the framework is explained in more detail in \cite{Zickendraht1965, Ballot1980}. In order to test the validity of our model, we consider the mass spectrum of a $B_{c}$ ($c\bar{b}$) meson in Appendix~\ref{sec:Bc}, where our results are shown together with experimental data.

    \subsection{Strong decays to $D^{(*)} B^{(*)}$ mesons or $D^{(*)}_{s} B^{(*)}_{s}$ mesons} \mbox{}

    Now we consider the decays of a $ c \bar{b} g $ into $ D^{(*) } B^{(*)}  $ mesons, where the gluon splitting process ($ g \rightarrow q \bar{q}$ with a light quark $q$) inside a $  c \bar{b} g $ takes place, estimating the decay widths for the processes. We also consider $D^{(*)}_{s} B^{(*)}_{s}$ as long as the energy thresholds for these decay channels are open. In the following explanation about the relevant formulation, we show the case of $D^{(*)} B^{(*)}$ mesons for illustration purposes. It is straightforward to apply the similar procedure to $D^{(*)}_{s} B^{(*)}_{s}$ mesons. In the formalism in Refs.~\cite{Iddir1998, Iddir2001}, the wave functions for both initial and final hadron states were represented by Gaussian-type functions. In our case, on the other hand, the initial wave function for a hybrid meson is given as the solution to the three-body equations \eqref{eq:Schrodinger_equation_for_three_body_system}, and the final wave functions are expressed by the Gaussian-type ones. Under this model, the decay processes are formulated in a nonrelativistic manner; this can be suitable for heavy mesons. Also, for later convenience we emphasize that a subscript $ q $ stands for a light quark ($u, d,$ or $s $) in this subsection.

    We enter into the explanation by defining the relevant relative momenta: $ {\bf P }_{c \bar{q} } $ is the relative momentum between $c $ and $ \bar{q} $; similarly ${\bf P}_{ \bar{b} q} $ is the relative momentum between $ \bar{b} $ and $ q $, and $ {\bf P}_{q \bar{q} } $ is the relative momentum between $q$ and $ \bar{q} $. Alongside the relative momenta, $ {\bf p}_{q} $ and ${\bf p}_{\bar{q} }$ are the momenta of $q$ and $ \bar{q} $, respectively. Using these momenta, we express the relative momenta as follows:       
    \begin{align}
    	{\bf P}_{c \bar{q} } &= \frac{m_{\bar{q}} {\bf p}_{c} - m_{c} {\bf p}_{\bar{q}} }{m_{c} + m_{\bar{q}}}	 = 
    	\frac{m_{\bar{q} } {\bf P}_{f} }{m_{c} + m_{\bar{q} } } + {\bf P}_{q \bar{q} }  - \frac{1}{2} {\bf k}_{g} \; , 
    	\\
    	{\bf P}_{\bar{b} q } &= \frac{m_{q } {\bf p}_{\bar{b}} - m_{\bar{b}} {\bf p}_{q} }{m_{\bar{b}} + m_{q}}	 = 
    	- \frac{m_{q} {\bf P}_{f} }{m_{\bar{b}} + m_{q } } - {\bf P}_{q \bar{q} }  - \frac{1}{2} {\bf k}_{g} \; , 
    	\\
    	{\bf P}_{q \bar{q} }   &=  \frac{{\bf p}_{q} - {\bf p}_{\bar{q} }  }{2}    \; , 
    \end{align} 	
    where $ {\bf P}_{f} = {\bf p}_{c} + {\bf p}_{\bar{q} } (= -{\bf p}_{\bar{b}} - {\bf p}_{q})$ is the final momentum of a $ c \bar{q} $ meson. 
    
    Next, in terms of the decay of a hybrid (signified by a subscript $H$) into $D^{(*)}$ and $B^{(*)}$ mesons, the formula of the decay width is given by  
    \begin{align}
    	\Gamma_{ H \rightarrow D^{(*) } B^{(* )}  } 
    	= 4 \alpha_{s} |  f_{H \rightarrow D^{(*) } B^{(* )} } |^{2} \frac{ |{\bf P}_{f }|   E_{ D^{(*)} } E_{ B^{(*)} }  }{ m_{H} }
    	\; ,
    \label{eq:decay_width_formula}
    \end{align}
    where $ E_{D^{(*)}} $ and $E_{B^{(*)}} $ are the energies of $D^{(*)}$ and $B^{(*)}$ mesons, respectively. $m_{H} $ is the mass of a hybrid meson. We also mention that $ f_{H \rightarrow D^{(*) } B^{(* )} } $ is the decay amplitude. The amplitude is written as
    \begin{align}
    	f_{H \rightarrow D^{(*) } B^{(* )} } = 
    	\sum \Omega \phi X I W 
    	\; , 
    \end{align}   
    where $ \Omega $, $\phi $, $X $, $I $, and $ W $ are the color part, the isospin part, the spin part, the overlap function, and the Clebsch-Gordan part, respectively \cite{Iddir1998}. The overlap function is defined by 
    \begin{align}
	I = \int   \frac{ d {\bf P}_{q \bar{q} } d {\bf k}_{g}  d \Omega_{f} }{ \sqrt{2 \omega }  (2 \pi)^{6} }
	\tilde{\Psi} ( {\bf P}_{f} + {\bf P}_{q \bar{q} }, {\bf k}_{g} ) 
	\psi_{D^{(*)}}^{*}({\bf P}_{c \bar{q}}) \psi_{B^{(*)}}^{*}({\bf P}_{\bar{b}q})
	Y_{LM}^{*} (\Omega_{f} ) \; ,  
	\label{eq:overlap_function_I_momentum_space}
    \end{align}	 
    where $ \Omega_{f} $ is the angular part of $ {\bf P}_{f} $. $\omega $ is the energy of the constituent gluon; and in Sec. \ref{sec:numerical_calculation_decay_widths}, we explain how to determine the value of $\omega $. Here $ \tilde{\Psi} $ is the momentum space representation of the initial wave function $\Psi  $ derived from solving the Eq. \eqref{eq:Schrodinger_equation_for_three_body_system}. $ \psi_{D^{(*)}}$ and $\psi_{B^{(*)}}$ are the wave functions of $D^{(*)} $ and $B^{(*)} $ mesons, respectively. As mentioned earlier, they are Gaussian-type functions:        
    \begin{align}
    	\psi_{D^{(*)}} ( {\bf P}_{c \bar{q} } ) & = \sqrt{ \frac{16\pi^{3} R_{c \bar{q} }^{2 L_{D^{(*)}} +3 }  }{\Gamma (\frac{3}{2} + L_{D^{(*)}} ) } } P_{c \bar{q} }^{ L_{D^{(*)}} } Y_{L_{D^{(*)}} M_{L_{D^{(*)}} } } e^{ -  \frac{1}{2} ( R_{c \bar{q} } P_{c \bar{q} } )^{2 }  } \; 
    	,
    	\nonumber  \\ 
    	\psi_{B^{(*)}} ( {\bf P}_{\bar{b} q } ) & = \sqrt{ \frac{16\pi^{3} R_{\bar{b} q}^{2 L_{B^{(*)}} +3 }  }{\Gamma (\frac{3}{2} + L_{B^{(*)}} ) } } P_{\bar{b} q}^{ L_{B^{(*)}} } Y_{L_{B^{(*)}} M_{L_{B^{(*)}} } } e^{ - \frac{1}{2} ( R_{\bar{b} q} P_{\bar{b} q})^{2 }  }
    	\; ,
    \end{align}    
    where $L_{D^{(*)}}$ and $L_{B^{(*)}}$ are the orbital angular momenta of $D^{(*)}$ and $B^{(*)}$ mesons, respectively. Similarly, $M_{L_{D^{(*)}}}$ and $M_{L_{B^{(*)}}}$ are their projections onto the $z$-axis. Here, $R_{c \bar{q} }$ and $ R_{\bar{b} q} $ are the Gaussian-type function's length parameters for the distance between $c $ and $\bar{q} $ in $D^{(*)}$ and the distance between $\bar{b}$ and $q $ in $B^{(*)}$, respectively. Their values will be set in the next section.

     
    After considering the above setup, we focus on the calculation of the decay amplitude in Eq.~(\ref{eq:decay_width_formula}); and to do so, we need to calculate the overlap function \eqref{eq:overlap_function_I_momentum_space} which contains multiple-integral for the momenta. Fortunately, we can decrease the dimension of the integral by going into detail about the exponential part of the function and integrating analytically with regard to the azimuthal angle $\phi_{g} $ of $ {\bf k}_{g} $. Hence, in the following, we describe the analytic expression of the integration with regard to $\phi_{g} $. First, the squares of $ {\bf P}_{c \bar{q} } $ and $ {\bf P}_{\bar{b} q}  $ are expressed by
    \begin{align}
    	{\bf P}_{c \bar{q} }^{2} & = \left( \frac{m_{\bar{q} }  {\bf P}_{f} }{m_{c} + m_{\bar{q} } } + {\bf P}_{q \bar{q} } \right)^{2} + \frac{1}{4} {\bf k}_{g}^{2}  
    	-  \frac{m_{\bar{q} }  {\bf P}_{f} \cdot {\bf k}_{g}  }{m_{c} + m_{\bar{q} } }   - {\bf P}_{q \bar{q} } \cdot {\bf k}_{g}  \;  
    	\nonumber \\
    	 & = \left( \frac{m_{\bar{q} }  {\bf P}_{f} }{m_{c} + m_{\bar{q} } } + {\bf P}_{q \bar{q} } \right)^{2} + \frac{1}{4} {\bf k}_{g}^{2}  
    	-  \frac{m_{\bar{q} }   P_{f} k_{g} \cos{ \theta_{g} }  }{m_{c} + m_{\bar{q} } }   -  P_{q \bar{q} } k_{g}  \cos{\theta_{q \bar{q} g} } \; ,    	
    	\\
    	{\bf P}_{\bar{b} q}^{2} & = \left( \frac{m_{q }  {\bf P}_{f} }{m_{\bar{b}} + m_{q} } + {\bf P}_{q \bar{q} } \right)^{2} + \frac{1}{4} {\bf k}_{g}^{2}  
    	+  \frac{m_{q }   P_{f} k_{g} \cos{ \theta_{g} }  }{m_{\bar{b}} + m_{q } }   +  P_{q \bar{q} } k_{g}  \cos{\theta_{q \bar{q} g} } \; ,
    \end{align} 
    where $ \theta_{q \bar{q} g} $ is the angle between $ {\bf P}_{q \bar{q} } $ and $ {\bf k}_{g} $. 
    Specifically, it is expressed as
    \begin{align}
    \cos{\theta_{q \bar{q} g} } = \sin{\theta_{q\bar{q} } } \cos{\phi_{q\bar{q}} } \sin{\theta_{g}} \cos{\phi_{g}}
    + \sin{\theta_{q \bar{q} } } \sin{\phi_{q\bar{q} }} \sin{\theta_{g}} \sin{\phi_{g} }
    + \cos{\theta_{q \bar{q} }} \cos{\theta_{g}} 
    \;  
    , 
    \end{align} 	
    where $(\theta_{q\bar{q}}$, $\phi_{q\bar{q}})$ and $( \theta_{g}, \phi_{g})$ are the angles in the spherical coordinates of ${\bf P}_{q \bar{q} }$ and ${\bf k}_{g}$, respectively.
    We note that here we choose the unit vector of $ {\bf P}_{f} $ as the zero degrees of the angular integration of $ {\bf k}_{g} $. Second, we perform the integration with regard to $\phi_{g} $. To do so, we write down the relevant part of the integrand: 
    \begin{align}
    	\exp \left( P_{q\bar{q} } k_{g}  \Bigg( \frac{R_{c \bar{q} }^{2} }{2 } - \frac{R_{\bar{b} q}^{2}  }{2}   \Bigg) \cos{ \theta_{q \bar{q} g} }  \right)  \; . 
    	\label{eq:integrand_exp_relevant_to_integration_with_phi_g}
    \end{align} 	
    Taking advantage of the formula relating to a modified Bessel function of the first kind $I_{0} $
    \begin{align}
      \int_{0}^{2 \pi} e^{ u \cos{\phi_{g}} + v \sin{\phi_{g}} } d \phi_{g}  =	2 \pi I_{0} (  \sqrt{u^{2} + v^{2} } )
    	\; ,  
    \end{align} 	
    where $u$ and $v$ are the functions parametrized as the coefficients of $\cos{\phi_{g}}$ and $\sin{\phi_{g}}$, respectively, in Eq.~\eqref{eq:integrand_exp_relevant_to_integration_with_phi_g}, we integrate with regard to $\phi_{g}$. This yields      
    \begin{align}
    \int_{0}^{2 \pi} d \phi_{g} 
    	\exp \left( P_{q\bar{q} } k_{g} \Bigg( \frac{R_{c \bar{q} }^{2} }{2 } - \frac{R_{\bar{b} q}^{2}  }{2}  \Bigg) \cos{ \theta_{q \bar{q} g} }  \right)
    	& =
    	2 \pi I_{0} \! \left( \left|  P_{q\bar{q} } k_{g} \Bigg( \frac{R_{c \bar{q} }^{2} }{2 } - \frac{R_{\bar{b} q}^{2}  }{2}  \Bigg)   \sin{\theta_{q \bar{q} } } \sin{\theta_{g} } \right|   \right)  
    	\nonumber  
    	\\ 
    	&  \quad \times
    	\exp \! \left( P_{q\bar{q} } k_{g} \Bigg( \frac{R_{c \bar{q} }^{2} }{2 } - \frac{R_{\bar{b} q}^{2}  }{2}  \Bigg)  \cos{\theta_{q \bar{q} } } \cos{\theta_{g} }  \right) 	
    	\; . 
    \end{align}

    As a result, the overlap function \eqref{eq:overlap_function_I_momentum_space} reduces to  
    \begin{align}
    		I & = \int   \frac{ d {\bf P}_{q \bar{q} } 
    			   k_{g}^{2} d k_{g}  d \cos{\theta_{g} }
    			  d \Omega_{f} }{ \sqrt{2 \omega }  (2 \pi)^{6} }
    	\tilde{\Psi} (  {\bf P}_{f} + {\bf P}_{q \bar{q} }, {\bf k}_{g} ) 
    	Y_{LM}^{*} (\Omega_{f } ) 
    	\nonumber  \\
    	&  \quad  \times     	
    	\sqrt{ \frac{16\pi^{3} R_{\bar{b} q}^{2 L_{B^{(*)}} +3 }  }{\Gamma (\frac{3}{2} + L_{B^{(*)}} ) } } P_{\bar{b}q}^{ L_{B^{(*)}} } Y_{L_{B^{(*)}} M_{L_{B^{(*)}} } }^{* }
    	\sqrt{ \frac{16\pi^{3} R_{c \bar{q}}^{2 L_{D^{(*)}} +3 }  }{\Gamma (\frac{3}{2} + L_{D^{(*)}} ) } } P_{c\bar{q}}^{ L_{D^{(*)}} } Y_{L_{D} M_{L_{D^{(*)}} } }^{* } 
    	\nonumber   \\
    	& \quad \times 
    	\exp \! \left(  - \frac{R_{c\bar{q} }^{2} }{2}  \left[  \left( \frac{m_{\bar{q} }  {\bf P}_{f} }{m_{c} + m_{\bar{q} } } + {\bf P}_{q \bar{q} } \right)^{2} + \frac{{\bf k}_{g}^{2}}{4}  
    	-  \frac{m_{\bar{q} }   P_{f} k_{g} \cos{ \theta_{g} }  }{m_{c} + m_{\bar{q} } }   \right]
    	 -  \frac{R_{\bar{b} q}^{2} }{2}
    	 \left[ \left( \frac{m_{q }  {\bf P}_{f} }{m_{\bar{b}} + m_{q} } + {\bf P}_{q \bar{q} } \right)^{2} + \frac{{\bf k}_{g}^{2}}{4}   
    	 +  \frac{m_{q }   P_{f} k_{g} \cos{ \theta_{g} }  }{m_{\bar{b}} + m_{q } } \right] 
    	  \right) 
    	\nonumber   \\
    	&  \quad \times 
    	2 \pi I_{0} \! \left( |  P_{q\bar{q} } k_{g} \Bigg( \frac{R_{c \bar{q} }^{2} }{2 } - \frac{R_{\bar{b} q}^{2}  }{2}  \Bigg)   \sin{\theta_{q \bar{q} } } \sin{\theta_{g} } |   \right)  
	\times
    	\exp \! \left( P_{q\bar{q} } k_{g} \Bigg( \frac{R_{c \bar{q} }^{2} }{2 } - \frac{R_{\bar{b} q}^{2}  }{2}  \Bigg)  \cos{\theta_{q \bar{q} } } \cos{\theta_{g} }  \right) 	
    	\; . 
    \end{align} 	
    Thus we see that the dimension of the original integration is decreased to seven. The above procedure freshly minted in the present research is helpful in reducing the numerical cost.

    \section{Numerical results}

    \subsection{Mass spectra of a $c \bar{b} g $}
    
    Now we conduct numerical calculations in terms of a $c\bar{b} g$. First of all, we consider the internal structure relating to the quantum number $J^{P}=1^{-} $ for the hybrid meson within the selection rules. The internal states of the hybrid meson are categorized according to the type of a constituent gluon \cite{Yaouanc1985}; it is either electric or magnetic. Table \ref{tab:numerical_results_low-lying_states_quantum_numbers} shows the possible low-lying states of a $ c \bar{b} g $, where E and M in the gluon column stand for the electric gluon and the magnetic gluon, respectively. We consider the lowest angular momenta only for each state. Strictly speaking, we also need to take account of higher states which are coupled to the low-lying states. The present procedure is a good approximation, because higher states should be suppressed as far as the states with lower energy are concerned.       
    \begin{table}[h]
    	\centering
    	\caption{Low-lying states for a $ c \bar{b} g $. In the first column, E and M mean the electric gluon and magnetic gluon, respectively. In the second column, $L_{c \bar{b}}$ is the relative orbital angular momentum between $c$ and $\bar{b}$. In the third column, $L_{g}$ is the orbital angular momentum of $g$ relative to the center-of-mass of $c \bar{b} $. In the fourth column, $ L_{tot} $ is the sum of $L_{c \bar{b} }$ and $L_{g} $. Meanwhile, $S_{c \bar{b} }$, $S_{g}$, and $S_{tot}$ are the total spin of $c$ and $\bar{b}$, the spin of $g$, and the total spin, respectively. Also, $J_{c\bar{b}}^{PC}$ is the total angular momentum of the $c$-$\bar{b}$ subsystem, and $J_{g}^{P}$ is the total angular momentum of $g$.} 
    	\label{tab:numerical_results_low-lying_states_quantum_numbers}	
    	\begin{tabular}{llllllllll}    
    	\hline \hline \\ [-0.3cm]	 
    		Gluon &  $ L_{ c \bar{b} } $ &   $ L_{g} $  & $ L_{tot}  $  
    		&  $S_{ c  \bar{b} }$  &   $S_{g}$   &  $S_{tot}$  & $ J_{ c \bar{b} }^{P} $ & $J_{g}^{P}$ & $ J^{P} $  \\
    		\hline 
    		E	  &    $1 $  &  $0 $  &   $1 $  &  $1 $  &   $1$   &  $0,1,2 $ & $ (0,1,2)^{+} $ & $1^{-}$ & $1^{-}$ \\  
    		M	  &    $0 $  &  $1 $  &   $1 $  &  $0 $  &   $1$   &  $1 $  & $ 0^{-}$ & $ 1^{+} $ &  $1^{-}$ \\ 		
    		\hline \hline 
    	\end{tabular}  
    \end{table}
  
   Then, we determine the masses and the wave functions of the hybrid mesons with magnetic and electric gluon in the Hamiltonian \eqref{eq:Schrodinger_equation_for_three_body_system} by using the auxiliary field method. The relevant parameters in the Hamiltonian are summarized in Table \ref{tab:spectrum:charmonium_hybrid:relevant_parameters}; these parameter sets are taken from earlier studies where the relevant parameters were fitted into the experimental data about the mass spectra of a charmonium ($c\bar{c}$) as well as a bottomonium ($b\bar{b}$) \cite{Kalashnikova2008, Mathieu2009, Miyamoto2018}. We also carry out the calculations in terms of the mass spectrum of a $B_{c}$ meson and confirm that the present framework provides a relatively fair description of the particle, as is seen in Appendix \ref{sec:Bc}.    

\begin{figure}[h]
	\centering
	\includegraphics[scale=0.45]{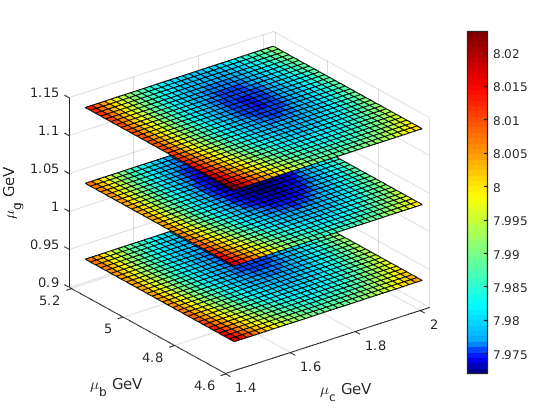} 
	\includegraphics[scale=0.45]{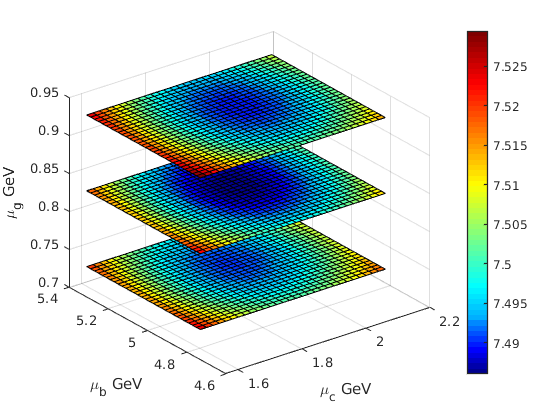} 
	\caption{The demonstration of the variational calculation of a hybrid meson $c\bar{b} g$ as functions of $\mu_{c}$, $\mu_{b} $, and $\mu_{g} $. The left graph is for the magnetic gluon case, whereas the right is for the electric case. The unit of the energies in the bars is GeV.}  
	\label{fig:auxiliary_field_method_cbbarg_mb5_magnetic_and_electric}
    \end{figure}

   Within the auxiliary field framework, we perform the variational calculations to minimize the energy of a $c \bar{b} g$ by changing $\mu_{c}$, $\mu_{\bar{b}}$, and $\mu_{g} $. We show the results of the calculations in Fig. \ref{fig:auxiliary_field_method_cbbarg_mb5_magnetic_and_electric}, where we can see that the local minimum values appear. We find that the lowest state is the electric gluon hybrid with a mass of about 7.48 GeV, which is in line with earlier research \cite{Akbar2018}.
    \begin{table}[h]
	\centering 	
	\caption{The parameters relating to the mass spectra of a hybrid meson $c \bar{b} g $. E and M in the gluon column of the table mean the electric gluon case and magnetic gluon case, respectively. Their masses are shown together.}
	\begin{tabular}{lllll lllll l}  
		\hline \hline  \\ [-0.3cm]
		& $ \sigma $  & $\alpha_{s} $  &    $ m_{c} $ & $m_{\bar{b}}$ &  $ J^{P} $ & Gluon & $ \mu_{c} $ & $ \mu_{b} $ & $ \mu_{g} $ & Mass   \\ 
		& ($ {\rm GeV}^{2} $) &      &     (GeV)  &  (GeV) & &  & (GeV)   &  (GeV)  & (GeV) & (GeV)   \\ 
		\hline  \\ [-0.3cm]
		& 0.16 & 0.55 & 1.48 & 5.00  & $ 1^{-} $  & M  & 1.77  & 4.99  & 1.03 & 7.97   \\ 
		& 0.16 & 0.55 & 1.48 & 5.00  & $ 1^{-} $  & E  & 1.86  & 5.05  & 0.84 & 7.48   \\ 
		\hline \hline 
	\end{tabular}
	\label{tab:spectrum:charmonium_hybrid:relevant_parameters}
    \end{table}

\begin{figure}
	\centering
	\includegraphics[scale=0.4]{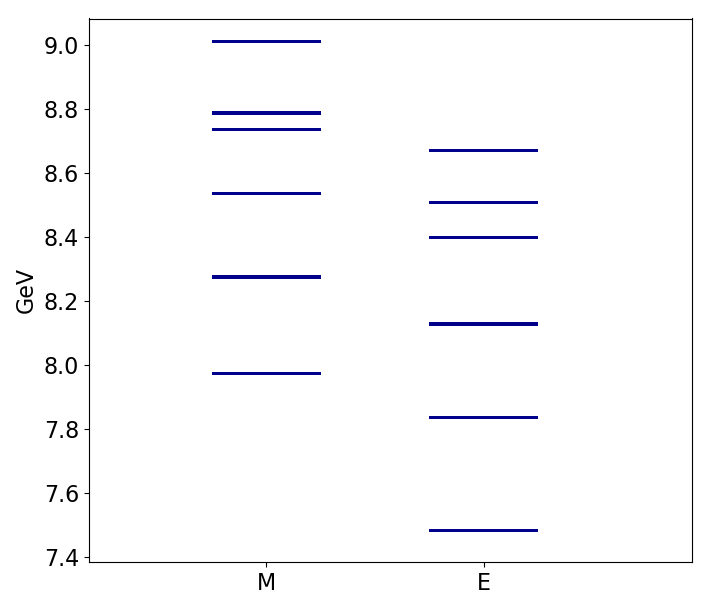}
	\caption{The mass spectra of the hybrid meson $c \bar{b} g$ with a magnetic gluon (M) and an electric gluon (E).}
	\label{fig:spectra_cbbarg_mb5}
    \end{figure}

    After calculating the masses for the lowest and the excited states, we show the mass spectra of magnetic and electric $ c \bar{b} g $ hybrids in Fig. \ref{fig:spectra_cbbarg_mb5}. We see from the figure that the states of a magnetic gluon hybrid are generally higher than those of an electric gluon hybrid, and that both of them appear above the $ DB  $ threshold ($  \sim  7.15 $ GeV). This result is consistent with earlier studies where the masses of a $ c \bar{c} g $ and a $ b \bar{b} g $ were obtained \cite{Mathieu2009, Miyamoto2018}.    

    \subsection{Decay widths} \label{sec:numerical_calculation_decay_widths} 
    We then move on to estimate the decay widths for the processes of a $ c \bar{b} g$ into $ D^{(*)} B^{(*)} $ and $D_{s}^{(*)} B_{s}^{(*)}$. We consider the electric gluon only, because the hybrid meson with a magnetic gluon does not decay into $D^{(*)} B^{(*)} $ and $ D_{s}^{(*)} B_{s}^{(*)} $ \cite{Iddir1998, Kou2005}.

     The values of the relevant parameters, $ \omega_{i} $, $ \mu_{i}$, and $R_{i}$ ($i=D^{(*)},B^{(*)}$ or $D_{s}^{(*)}, B_{s}^{(*)}$) are summarized in Table \ref{tab:hybrid_decay_width_mesons_parameters}, where the values in parentheses mean the values for the cases where a strange quark involves. We note that here $ \mu_{i} $ is the reduced mass of a $ D $ or a $ B $ meson. They are expressed as      
	$ \mu_{D} = \mu_{D^{*}}= \frac{m_{c} m_{ \bar{q} } }{m_{c} +  m_{\bar{q}  } } $ and   
	$ \mu_{B} = \mu_{B^{*}}= \frac{m_{\bar{b} } m_{ q } }{m_{ \bar{b} } +  m_{ q  } } $, where the reduced mass (as a \texttt{"}static\texttt{"} variable) for $D$ is identical to that of $D^{*}$, and similarly for $B$ and $B^{*}$. In the present study, we have $\mu_{D}=0.283$ GeV ($\mu_{D_{s}}=0.373$ GeV) and $ \mu_{B} =0.327$ GeV ($\mu_{B_{s}}=0.454$ GeV), because we choose 350 MeV for the mass of $ u $ and $d $ quarks and 500 MeV for an $s$ quark \cite{Iddir1998, Iddir2001}. When it comes to $\omega_{D}$, it is the roughly averaged energy difference between a $D$ meson's ground state and its first excited state ($D_{0}(2400)^{0}$ and $ D_{s1}(2460)^{\pm}$); and for $\omega_{B}$, the first excited state is $B_{1}(5721)$\footnote{For instance, $w_{D} = ( M(D_{0}(2400)) - M(D^{\pm}) + M(D_{s1}(2460)) - M(D_{s}) ) / 2$, where $M$ means the mass of a particle. Also, we choose $B_{1}$ not $B_{0}$ as the first excited state, because the latter has not been discovered yet \cite{Tanabashi2018}.}. In addition, $R_{i}$ is connected to these variables via $R_{i}^{2} = \frac{1}{\omega_{i} \mu_{i}}$ \cite{Iddir1998}. In parallel with determining these variables, we need to estimate the gluon energy $ \omega $ in the overlap function \eqref{eq:overlap_function_I_momentum_space}. Recognizing that basically this is related to the effective mass of a gluon in the auxiliary field method, we use the effective gluon mass $ \mu_{g} (= 0.84  \; \; \textrm{GeV} )$ for $\omega$ in the present calculation\footnote{This approach has not been adopted in earlier research. In the previous studies \cite{Iddir1998, Iddir2001}, including ours \cite{Miyamoto2018}, the value of $    \omega $ was determined to be practically identical to the frequency of the harmonic confinement potential by which the quarks and gluons are confined as the hybrid mesons.}.

      \begin{center} 	  
      \begin{table}[h]  	 	
   	  \caption{The relevant parameters for the decay processes of hybrid mesons into $D^{(*)}B^{(*)}$ ($D^{(*)}_{s}B^{(*)}_{s}$) mesons. $m_{c} $ and $ m_{\bar{b}} $ are the masses of a charm quark and an antibottom quark, respectively. The values in parentheses stand for the values for the cases where a strange quark is involved. Here, we note that $\omega_{D} = \omega_{D_{s}}$ and $\omega_{B}=\omega_{B_{s}}$.}
   	  \begin{tabular}{ll P{2.0cm} P{2.0cm} P{3.5cm} P{4.0cm} ll  }    
   	  \hline \hline \\ [-0.3cm]	 
   		&  & $ m_{c} $ or $ m_{\bar{b}} $ & $ \omega_{D} $ or $\omega_{B}$ & $ \mu_{D} $ ($\mu_{D_{s}}$) or $ \mu_{B }$ ($\mu_{B_{s}}$)  &  $ R_{D} $ ($R_{D_{s}}$) or $R_{B}$ ($R_{B_{s}}$)  &    &         \\ 
   		& &     (GeV)    &  (GeV) &   (GeV)    &  (${\rm GeV^{-1} } $)  &          & \\
   		\hline \\ [-0.3cm]   
   		\textit{D} meson & & 1.48   &  0.468     & 0.283 (0.373)  &  2.74 (2.39)  &  &      \\ 
   		\textit{B} meson & & 5.00 &    0.446     &   0.327 (0.454) &  2.61 (2.22)   & &        \\ 
   		\hline \hline 
   	  \end{tabular}
   	  \label{tab:hybrid_decay_width_mesons_parameters}
   	  \end{table} 
      \end{center}

     \begin{center}
     	\begin{table}[h] 
     		\caption{The partial decay widths for $D^{(*)} B^{(*)}$ channels from the hybrid meson $c\bar{b}g$ with an electric gluon. Here $ D^{* 0} = D^{*}(2007)^{0} $ and $D^{* \pm} = D^{*} (2010)^{\pm } $. The mass of $u$ and $d$ quarks is 350 MeV, while the mass of an $s$ quark is 500 MeV. Here, $ DB = D^{+} B^{0} + D^{0} B^{+} $, $ D^{*}B = D^{*+} B^{0} + D^{*0} B^{+} $, $ DB^{*} = D^{+} B^{*0} + D^{0} B^{*+} $ and $ D^{*}B^{*} = D^{*+} B^{*0} + D^{*0} B^{*+} $. The unit of the decay widths in the table is MeV.}
     		\begin{tabular}{lll lll lll lll l}
     			\hline 
     			\hline  \\ [-0.3cm]
     			Channel       & 	 & $ DB $ &  & $ D^{*} B$ &   & $D B^{*} $ &  &  $ D^{*} B^{*} $ &    &  $ D_{s}^{+} B_{s}^{0} $ & $ D_{s}^{+} B_{s}^{*} $  &  Total \\
     			\hline \\ [-0.3cm]
     			$ \Gamma_{\scriptscriptstyle GS \rightarrow D^{(*)} B^{(*)}  \text{or} D_{s}^{(*)} B_{s}^{(*)} } $      & $J_{c \bar{b}} $	&  & & & & & & &  & &  & \\ 
     			&0  & 74.56 & & 114.01 & & 140.07 & & 313.37  & &   4.19 &  5.29 & 651.5 \\ 
     			&1 & 223.72 & &  85.50 & & 105.04 & &  268.59 & &  12.58 &   3.97 & 699.4  \\ 
     			&2 & 372.87 & & 142.51 & & 175.08 & &  89.52  & &  20.98 &   6.62 & 807.6 \\ 
     			     			     			
     			$ \Gamma_{\scriptscriptstyle 1st \rightarrow D^{(*)}  B^{(*)} \text{or} D_{s}^{(*)} B_{s}^{(*)} } $      & $J_{c \bar{b}} $	&  & & & & & & &  & & & \\
     			&0 & 18.68 & &  28.76 & &  34.31 & &  84.52 & &   1.01 &   1.51 & 168.8 \\ 
     			&1 & 56.05 & &  21.57 & &  25.73 & &  72.44 & &   3.05 &   1.14 & 180.0 \\ 
     			&2 & 93.44 & &  35.97 & &  42.89 & &  24.14 & &   5.09 &   1.89 & 203.5 \\ 
     			\hline \hline 
     		\end{tabular}
     		\label{tab:decay_into_DB_widths_values}
     	\end{table}
     \end{center}

     Now, we calculate the decay widths for relevant channels. Using shorthand notations, we consider the following channels: $ DB = D^{+} B^{0} + D^{0} B^{+} $, $ D^{*}B = D^{*+} B^{0} + D^{*0} B^{+} $, $ DB^{*} = D^{+} B^{*0} + D^{0} B^{*+} $, $ D^{*}B^{*} = D^{*+} B^{*0} + D^{*0} B^{*+} $, $D_{s}^{+} B_{s}^{0}$, and $D_{s}^{+} B_{s}^{*}$. Those channels are allowed as the open thresholds.
     The calculated partial decay widths are shown in Table \ref{tab:decay_into_DB_widths_values}. As to the decays from the lowest state of a $c \bar{b} g$, the calculated widths exceed 0.6 GeV for $J_{c \bar{b} }= 0$, $1$, and $2$. This suggests that it seems unlikely that this state can be experimentally confirmed. When it comes to the decays from the first excited state of a $c \bar{b} g$, the calculated widths are in the range of 150 MeV to 250 MeV. It seems that it could still be difficult to detect the decay signals for these channels; but there is a fighting chance that a $ c \bar{b} g$ will be observed as its first excited state. 
     
     Also, it turns out that the decay widths have a heavy dependence on the choice of the final states, $DB$, $D^{*}B$, $DB^{*}$, $D^{*}B^{*}$, $D_{s}^{+} B_{s}^{0}$, and $D_{s}^{+} B_{s}^{*}$. The differences, which arise partly from the combinatorial factor of the internal spins and the limitation of the available phase space, should be reflected in the branching ratios when they are experimentally measured in the future. Hence, we expect that the validity of our hybrid model will be assessed by further experiments, on the basis of not only the absolute values of the decay widths but also the branching ratios.

	 It is worth mentioning the uncertainty of the obtained decay widths. In the present study, we set $ \alpha_{s} = 0.55 $ to obtain the mass spectra of a $c \bar{b} g$; we use the same $\alpha_{s} $ in calculating the decay widths. However, actual $\alpha_{s}  $ for the decays into $ D^{(*) } B^{(*)} $ or $D_{s}^{(*)} B_{s}^{(*)} $ can be smaller than $0.55$. This fact allows us to argue that actual decay widths can be smaller than the values which are presented in Table \ref{tab:decay_into_DB_widths_values}. We comment that the branching ratios are unaffected by the uncertainty of $\alpha_{s}$ in the present framework.

	
	\section{Conclusion}
    We study a bottom-charm hybrid meson $c \bar{b } g$ in the constituent gluon model, considering its mass spectra and decays into $ D^{(*)} B^{(*)} $ mesons and $ D^{(*)}_{s} B^{(*)}_{s} $ mesons. 
    We obtain the mass spectra for magnetic gluon and electric gluon hybrids. For a magnetic gluon hybrid, the lowest state appears at around 8 GeV, while for an electric gluon hybrid, the lowest state appears at around 7.5 GeV. We find that the states for a magnetic gluon hybrid are higher than those for an electric gluon hybrid and that both of them sit above the $DB $ threshold. These results are in line with earlier research.           
    Also, we estimate the decay widths of the relevant $ D^{(*) } B^{(*)} $ and $ D^{(*)}_{s} B^{(*)}_{s} $ channels. For the decays of the lowest state of a $ c \bar{b} g $, the obtained widths seem large to the extent that it could be experimentally difficult to confirm the state. For those of the first excited state, although the calculated widths are still large, there could be a prospect that a $ c \bar{b}   g $ will be detected as its first excited state in the future.
    
    As part of future work, we mention that higher multichannel effects for the spectra of a hybrid and loop effects for the decays will be studied. We also mention that the other quantum numbers such as $ J^{P} = 0^{ \pm } $ will be considered. Exploring the possibility that there could exist other types of hybrid mesons such as $ s \bar{c} g $ and $ s \bar{b} g $ might generate discussions about flavor dependence in the hybrid mesons. Moreover, it will be worthwhile to consider gluonic excitations not just in heavy mesons but in heavy baryons such as $ \Omega_{ccc}^{++} $. We hope that experimental research projects will be set up in order to search for these hybrid mesons and baryons and that related experiments will be carried out at high energy beam facilities such as the LHC in the future.                
    
    \section*{Acknowledgments}
    This work is supported by the Grant-in-Aid for Scientific Research (Grants No. 17K05435) from Japan Society for the Promotion of Science and by the MEXT-Supported Program for the Strategic Foundation at Private Universities, \texttt{"}Topological Science\texttt{"} under Grant No. S1511006.

	\appendix
	\section{Mass spectrum of a $B_{c}$ meson} \label{sec:Bc}
	We assess the validity of our model by examining whether our model can reproduce the experimental data on the mass spectrum of a $ B_{c} $ $ (c \bar{b} )$. Accordingly we calculate the mass spectrum of a $B_{c} $ as a two-body system of $c$ and $\bar{b}$, treating the particle on the same footing as a $ c \bar{b} g $: we use the Salpeter-type free Hamiltonian and the auxiliary field method. As stated in Section \ref{sec:Hamiltonian_and_auxiliary_field_method}, we therefore use the auxiliary fields $\nu_{c} $ and $\nu_{b} $ for $c$ and $\bar{b} $, respectively, in order to reexpress the whole Hamiltonian as 
	\begin{align}
		H_{c\bar{b}} & = \sqrt{{\bf p}_{c}^{2} + m_{c}^{2}  } + \sqrt{{\bf p}_{\bar{b}}^{2} + m_{b}^{2}  }  
		+ V_{c \bar{b}}  \\
		& =  \frac{\nu_{c} + \nu_{b} }{2} + \frac{m_{c}^{2}}{2\nu_{c} } + \frac{m_{b}^{2} }{2\nu_{b} } 
		 + \frac{{\bf p}_{c}^{2} }{2 \nu_{c} } + \frac{{\bf p}_{\bar{b}}^{2}}{2 \nu_{b} } 
		 + V_{c\bar{b}}
		 \; , 
	\end{align}
    where $V_{c\bar{b}}$ is an inter-quark potential whose explicit form is given later. As we now consider the center-of-mass frame, we have $ {\bf p}_{c} = - {\bf p}_{\bar{b} } = {\bf p} $. Substituting ${\bf p} $ into the above Hamiltonian, we obtain the following simple expression:
    \begin{align}
    H_{c\bar{b}}	 =  \frac{\nu_{c} + \nu_{b} }{2} + \frac{m_{c}^{2}}{2\nu_{c} } + \frac{m_{b}^{2} }{2\nu_{b} } 
    	+ \frac{{\bf p}^{2} }{2 \nu } 
    	+ V_{c\bar{b}}  \; , 
    \end{align} 	    
	where 
	\begin{align}
		\nu = \frac{\nu_{c} \nu_{b}}{\nu_{c} + \nu_{b} } \; . 
	\end{align} 	
    When it comes to the interaction between charm and anti-bottom quarks, we use 
    \begin{align}
    	 V_{c\bar{b}} = V_{L+C}^{c\bar{b}} + V_{SS}^{c\bar{b}} + V_{T}^{c\bar{b}} \; ,
    \end{align} 	
    where $V_{L+C}^{c\bar{b}} $, $V_{SS}^{c\bar{b}}$ and $V_{T }^{c\bar{b}}$ are the linear and color-Coulomb potentials, the spin-spin term and the tensor interaction, respectively. In this case, we focus on $ ^{1}S_{0} $ and $ ^{3}S_{1} $; there is no need of spin-orbit interactions. We now express the potentials explicitly as follows:     
    \begin{align}
      V_{L+C}^{c\bar{b}} & =  \sigma r_{c \bar{b} }  + \left( - \frac{4}{3} \right) \frac{ \alpha_{s} }{ r_{c \bar{b} } }  \; ,   \\
      V_{SS}^{c\bar{b}} & =  \left(  - \frac{4}{3}  \right) \frac{ (-8) \pi \alpha_{s} }{3 \mu_{ c} \mu_{\bar{b} } } ({\bf s}_{c} \cdot {\bf s}_{\bar{b}} ) \delta ({\bf r}_{c \bar{b} })   \; , \\ 
      V_{T}^{c\bar{b}} & =  \frac{4 \alpha_{s } }{3 \mu_{c} \mu_{\bar{b} } r_{c\bar{b} }^{5} }  \left[ 3 ({\bf s}_{c} \cdot {\bf r}_{c\bar{b} }  ) ({\bf s}_{\bar{b}} \cdot  {\bf r}_{c \bar{b}} ) - r^{2} ({\bf s}_{c} \cdot {\bf s}_{\bar{b} } )  \right]   \; ,  	
    \end{align}   
    where we set $ \sigma=0.16 \; \textrm{GeV}^{2} $ and $\alpha_{s}=0.55 $ as we used them for a $c\bar{b}g$. We note that we do not consider the S-D mixing, and thus the tensor term does not affect the results. When it comes to the $V_{SS}$, we substitute the following exponential function for the delta function of the spin-spin term \cite{Yoshida2015}:
    \begin{align}
    \delta ({\bf r} ) \rightarrow  	\frac{\Lambda^{2} }{4 \pi r} e^{ - \Lambda r }  \; ,  
    \end{align} 	
    where $\Lambda $ is set to $3.5 \; {\rm fm}^{-1}$ \cite{Yoshida2015}. We set $m_{c}=1.48 $ GeV and $ m_{\bar{b}}=5.00$ GeV as the same values in the calculation for a $c\bar{b}g$. These values are fitted to the mass spectra of a charmonium and a bottomonium; for further fittings, fine-tuning can be necessary.

\begin{figure}[h]
    	\centering     	
    	\includegraphics[scale=0.47]{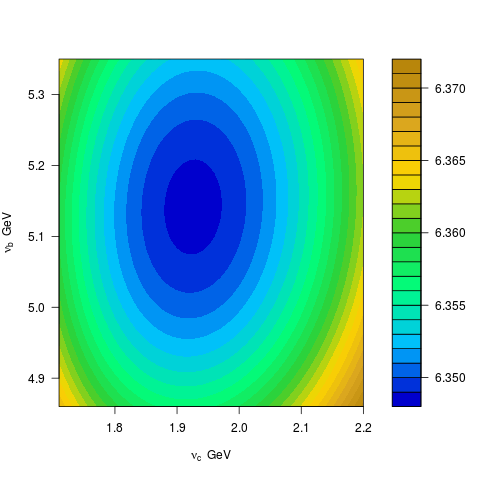}    	
    	\caption{The ground state energies of a $ c \bar{b} g $ meson as functions of $ \nu_{c} $ and $ \nu_{b} $. $\Lambda=3.5 \; {\rm fm}^{-1} $. The parameter settings are as follows: $   m_{c} =1.48 $ GeV, $m_{\bar{b}}=5.00$ GeV, $\alpha_{s}=0.55$, and $ \sigma=0.16 {\rm GeV}^{2}$. The unit of the values for the colorbar is GeV. We also mention that $ M_{B_{c} } =  6274.9 \pm 0.8$ MeV \cite{Tanabashi2018}.}
    	\label{fig:B_c_meson_auxiliary_field_method_ground_state}
    \end{figure}

    At these parameter settings, we calculate the ground state of a $B_{c} $. The result of the variational calculations is shown in Figure \ref{fig:B_c_meson_auxiliary_field_method_ground_state}, where we see that the ground state ($ \sim 6.34 $ GeV) exists (at $\nu_{c}=1.93$ GeV and $\nu_{b}=5.14$ GeV).  
          The calculated ground state energy fairly agrees with the experimental data ($M_{B_{c} } \sim 6.27 $ GeV) \cite{Tanabashi2018}. Meanwhile, the calculated energy of the $2S$ state is about 7.01 GeV; this appears about 0.17 GeV above the experimental data ($M_{B_{c}}(2S) \sim $ 6.84 GeV). 
           Thus, we confirm that the Salpeter-type Hamiltonian and the auxiliary field method provide a relatively fair description of a $B_{c}$ meson.


\begin{thebibliography}{70}
  	
  	
    \bibitem{Choi2003}
  	S.-K. Choi \textit{et al}., Phys. Rev. Lett. \textbf{91}, 262001 (2003).
	
	\bibitem{Tanabashi2018}
	M. Tanabashi et al. (Particle Data Group), Phys. Rev. D \textbf{98}, 030001 (2018).
	
    \bibitem{Aubert2005}
	B. Aubert \textit{et al}., Phys. Rev. Lett. \textbf{95}, 142001 (2005).	
	
	\bibitem{Dubynskiy2008}
	S. Dubynskiy and M. B. Voloshin, Phys. Lett. B \textbf{666}, 344 (2008).  
	
	
	\bibitem{Wang2013} 
	Q. Wang, C. Hanhart, and Q. Zhao, Phys. Rev. Lett. \textbf{111}, 132003 (2013).
	
	\bibitem{Maiani2005_Y4260}
	L. Maiani, F. Piccinini, A. D. Polosa, and V. Riquer, Phys. Rev. D \textbf{72}, 031502R (2005)   
	
	\bibitem{Kou2005}
	E. Kou and O. P\`{e}ne, Phys. Lett. B \textbf{631}, 164 (2005).
	
	
	\bibitem{Olsen2018} 
	S. L. Olsen, T. Skwarnicki, and D. Zieminska, Rev. Mod. Phys. \textbf{90}, 015003 (2018)
	
	
	
	\bibitem{Ding2007}
	G.-J. Ding and M.-L. Yan, Phys. Lett. B \textbf{650}, 390 (2007).      
	
	\bibitem{Ding2008} 
	G.-J. Ding, J.-J. Zhu, and M.-L. Yan, Phys. Rev. D \textbf{77}, 014033 (2008).
	
		
	\bibitem{Isgur1985}
	N. Isgur and J. Paton, Phys. Rev. D, \textbf{31}, 2910 (1985).  
	
	
	\bibitem{Barnes1979}
	T. Barnes, Nucl. Phys. B \textbf{152}, 171 (1979). 
	 
	
	\bibitem{Giles1976}
	R. C. Giles and S.-H. H. Tye, Phys. Rev. Lett. \textbf{37}, 1175 (1976).    
	
	
	\bibitem{Horn1978}
	D. Horn and J. Mandula, Phys. Rev. D, \textbf{17}, 898 (1978).  
	
	
	
	\bibitem{Yaouanc1985} 
	A. Le Yaouanc, L. Oliver, O. P\`{e}ne, J.-C. Raynal, and S. Ono, Z. Phys. C \textbf{28}, 309 (1985).  
	

	\bibitem{Chen2014} 
	W. Chen, T. G. Steele and S.-L. Zhu, J. Phys. G: Nucl. Part. Phys. \textbf{41} (2014) 025003.  
		
	\bibitem{Akbar2018}
	N. Akbar, M. A. Sultan, B. Masud and F. Akram, arXiv:1811.07552 [hep-ph]
	
	
	

	
	
	
	
	
	
	
	
	
	

	
	
	
	

	
	
	
	
	
	
	
	
	
	
	
	
	
	
	
	
	
	
	
	
	
	
	
	
	
	
	
	
	
	\bibitem{Buisseret2006_hybrids_AF}
	F. Buisseret and V. Mathieu, Eur. Phys. J. A \textbf{29}, 343 (2006).
	
	\bibitem{Buisseret2006_effective_potential_hybrid}
	F. Buisseret and C. Semay, Phys. Rev. D \textbf{74}, 114018 (2006).
		
	\bibitem{Mathieu2009}
	V. Mathieu, Phys. Rev. D \textbf{80}, 014016 (2009).
	
	
	\bibitem{Kalashnikova2005}
	Yu. S. Kalashnikova and A. V. Nefediev, Phys. At. Nucl. \textbf{68}, 650 (2005). 
	
	
	
	
	
	\bibitem{Miyamoto2018}
	T. Miyamoto and S. Yasui, Phys. Rev. D \textbf{98}, 094027 (2018). 
	
	\bibitem{Zickendraht1965} 
	W. Zickendraht, Ann. Phys. (N.Y.) \textbf{35}, 18 (1965).
	
	\bibitem{Ballot1980}
	J. L. Ballot and M. F. De La Ripelle, Ann. Phys. (N.Y.) \textbf{127}, 62 (1980).
	
	
	
	
	
	
	\bibitem{Iddir1998}
	F. Iddir, S. Safir, and O. P\`{e}ne, Phys. Lett. B \textbf{433}, 125 (1998). 
	
	
	
	
	
	
	\bibitem{Iddir2001}
	F. Iddir and A. S. Safir, Phys. Lett. B \textbf{507}, 183 (2001).
	
	\bibitem{Kalashnikova2008}
	Yu. S. Kalashnikova and A. V. Nefediev, Phys. Rev. D \textbf{77}, 054025 (2008).	
    
		
	\bibitem{Yoshida2015}
	T. Yoshida, E. Hiyama, A. Hosaka, M. Oka, and K. Sadato, Phys. Rev. D \textbf{92}, 114029 (2015).
		
	
	
	
	
	
		
		
		
		
	
				
  \end{thebibliography}
\end{document}